\documentclass[aps,pra,twocolumn,amsmath,amssymb,superscriptaddress]{revtex4-1}
\usepackage{amsfonts}
\usepackage{amsmath}
\usepackage{amssymb}
\usepackage{graphicx}
\usepackage{afterpage}

\begin{document}
\title{Mid-Infrared ultra-high-Q resonators based on fluoride crystalline materials}

\author{C.~Lecaplain}
\thanks{These authors contributed equally to this work}
\affiliation{\'{E}cole Polytechnique F\'{e}d\'{e}rale de Lausanne (EPFL), CH-1015,
Lausanne, Switzerland}
\author{C.~Javerzac-Galy}
\thanks{These authors contributed equally to this work}
\affiliation{\'{E}cole Polytechnique F\'{e}d\'{e}rale de Lausanne (EPFL), CH-1015,
Lausanne, Switzerland}
\author{M. L.~Gorodetsky}
\affiliation{Russian Quantum Center, Skolkovo 143025, Russia}
\affiliation{Faculty of Physics, M. V. Lomonosov Moscow State University, Moscow 119991, Russia}
\author{T. J.~Kippenberg}
\email{tobias.kippenberg@epfl.ch}
\affiliation{\'{E}cole Polytechnique F\'{e}d\'{e}rale de Lausanne (EPFL), CH-1015,
Lausanne, Switzerland}

\begin{abstract}
\textbf{Decades ago, the losses of glasses in the near-infrared (near-IR) were investigated in views of developments for optical telecommunications \cite{Lines1986}. Today, properties in the mid-infrared (mid-IR) are of interest for molecular spectroscopy applications  \cite{Thorpe2006,Coddington2008,Thorpe2008a}. In particular, high-sensitivity spectroscopic techniques based on high-finesse mid-IR cavities hold high promise for medical applications \cite{Paul2001,Brown2003,Thorpe2008a}. Due to exceptional purity and low losses, whispering gallery mode microresonators based on polished alkaline earth metal fluoride crystals (i.e the $\mathrm{XF_2}$ family, where X $=$ Ca, Mg, Ba, Sr,...) have attained ultra-high quality (Q) factor resonances (Q$>$10$^{8}$) in the near-IR and visible spectral ranges \cite{Savchenkov2004}.
Here we report for the first time ultra-high Q factors in the mid-IR using crystalline microresonators.
Using an uncoated chalcogenide (ChG) tapered fiber, light from a continuous wave quantum cascade laser (QCL) is efficiently coupled to several crystalline microresonators at 4.4 $\mu$m wavelength. We measure the optical Q factor of fluoride crystals in the mid-IR using cavity ringdown technique. We observe that $\mathrm{MgF_2}$ microresonators feature quality factors that are very close to the fundamental absorption limit, as caused by the crystal's multiphonon absorption (Q$\sim$10$^{7}$), in contrast to near-IR measurements far away from these fundamental limits. Due to lower multiphonon absorption in $\mathrm{BaF_2}$ and $\mathrm{SrF_2}$, we show that ultra-high quality factors of Q $\geqslant$ 1.4 $\times 10^{8}$ can be reached at 4.4 $\mu$m. This corresponds to an optical finesse of $\mathcal{F}>$ 4$\cdot$ 10$^{4}$, the highest value achieved for any type of mid-IR resonator to date, and a more than 10-fold improvement over the state-of-the-art \cite{Schwarzl2007,AlligoodDeprince2013,Foltynowicz2013}. Such compact ultra-high Q crystalline microresonators provide a route for narrow-linewidth frequency-stabilized QCL \cite{SicilianideCumis2016} or mid-IR Kerr comb generation.}
\end{abstract}
\maketitle

\section{Introduction}

The mid-infrared (mid-IR) spectral window ($\lambda \sim$ 2.5-20 $\mu$m) is a highly useful range for spectroscopy, chemical and biological sensing, materials science and industry as it includes strong rotational-vibrational absorption lines of many molecules as well as two atmospheric transmission windows of 3-5 $\mu$m and 8-13 $\mu$m. In this band, the absorption strengths of molecular transitions are typically 10 to 1000 times greater than those in the visible or near-infrared (near-IR), offering the potential to identify the presence of substances with extremely high sensitivity and selectivity, as required in trace-gas detection \cite{Brown2003,Coddington2008,Foltynowicz2013}, breath analysis \cite{Thorpe2006,Thorpe2008a} and pharmaceutical process monitoring. The invention of the quantum cascade laser (QCL) \cite{Faist1994} was a scientific and technological milestone in the development of mid-IR laser sources. Today, continuous wave QCL operate at room temperature, offer high output power (Watts) and are a commercial technology \cite{Yao2012}. These features lead to their wide adoption for spectroscopic applications in the mid-IR region. However their free running linewidth is subject to undesired noise that makes their frequency stabilization challenging \cite{Argence2015}. High finesse optical cavities have the potential for substantial reduction of laser noise \cite{Kessler2012,Liang2015} and have been recently investigated \cite{SicilianideCumis2016}.\\ 
The exploration of ultra-high quality (Q) cavities in the mid-IR is a relatively new area driven by the development of QCL frequency stabilization \cite{SicilianideCumis2016}, ultra-sensitive molecular sensors \cite{Vollmer2008}, cavity-based spectroscopy \cite{Thorpe2006,Orr2011} and optical frequency combs \cite{Kipp2011}. ZBLAN \cite{Way2012} and chalcogenide glasses \cite{Ma2015} resonators appear as promising for mid-IR wavelengths. Moreover, the silicon platform \cite{Singh2014} have enabled the fabrication of mid-IR microresonators at $\lambda\sim $ 5 $\mu$m. To date, Q factors of $10^{5}$ are already achieved in silicon-on-sapphire \cite{Shankar2013} and chalcogenide glass-on-silicon \cite{Lin2013}, whereas Q factors of $10^{4}$ are obtained in silicon-on-calcium fluoride \cite{Chen2014}. Yet to date, achieving ultra-high Q factors in the mid-IR (or correspondingly high optical finesse) is an outstanding challenge. In the near-IR, high finesse is achieved using Fabry-P\'{e}rot cavities based on supermirrors \cite{Cole2013} or using microcavities based on WGM. Interference coatings based on substrate-transferred single-crystal multilayers show the potential for $\leqslant$100 ppm of optical losses in the mid-infrared (potentially out to 7 $\mu$m) \cite{Cole2016}.\\

\begin{figure*}[ht]
	\centering
	\includegraphics[width=1\linewidth]{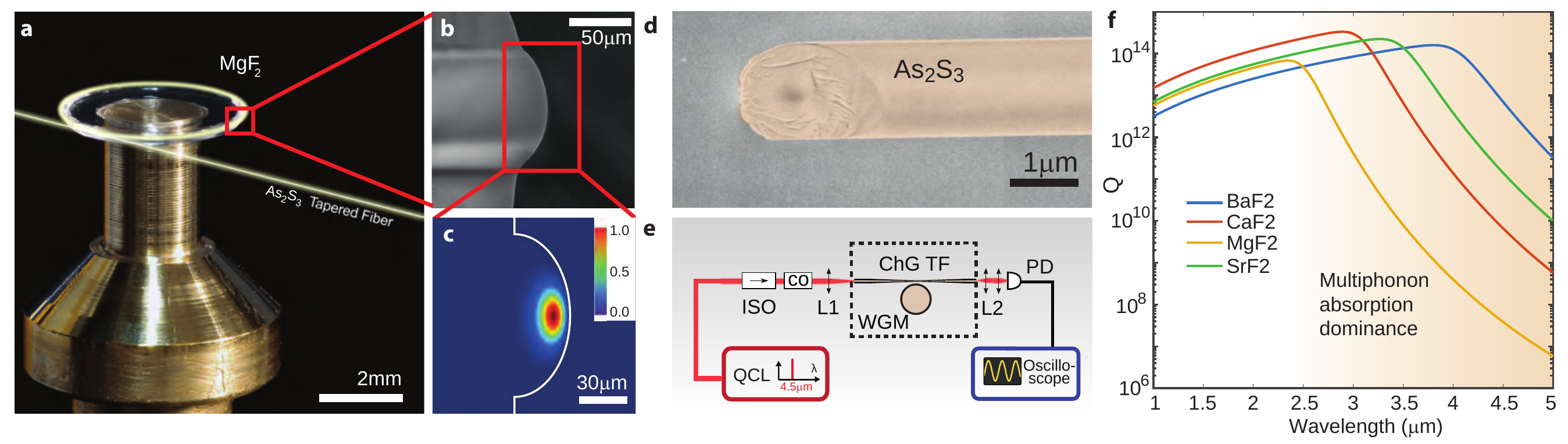}
	\caption{\textbf{Alkaline earth metal fluoride based crystalline microresonators, chalcogenide tapered fiber technology and Q factor dependence of fluoride materials. a} Magnesium fluoride (MgF$_{2}$) crystalline microresonator with a diameter of $\sim$5 mm. Whispering Gallery Modes (WGMs) of the microresonator are excited via evanescent coupling using a chalcogenide i.e. ChG (As$_{2}$S$_{3}$) tapered fiber. \textbf{b} Scanning electron microscope (SEM) image of the MgF$_{2}$ protrusion. Its radius of curvature, which confines the mode in the
		azimuthal direction, is $\sim$ 50  $\mu$m.  \textbf{c} Finite element model simulations of the optical intensity profile of the fundamental WGM at $\lambda$ = $4.5 \ \mu$m.\ \textbf{d} SEM image of the waist of a ChG tapered fiber with subwavelength diameter of 1.2 $\mu$m. \textbf{e} Experimental setup composed of a WGM microresonator pumped by a quantum cascade laser (QCL) evanescently coupled through a chalcogenide tapered fiber (ChG TF), followed by an oscilloscope to record the transmission. An optical isolator (ISO) protects the pump laser from Fresnel reflection ($\sim 14 \%$) at the cleaved fiber ends. Mid-IR free space control optics (CO), including waveplates, neutral densities and a mid-IR electro-optic modulator. L1, L2 are lenses for free space coupling into the ChG TF. PD, photodetector. Tapered fiber and microresonator are kept under a dry and inert atmosphere to preserve from degradation. \textbf{f} Quality factor dependence of different crystals with respect to the wavelength. For mid-IR wavelengths, multiphonon absorption competes with Rayleigh scattering and strongly impacts the Q factor. }
	\label{figure1}
\end{figure*}

Owing to low material losses, crystalline whispering gallery mode (WGM) microresonators \cite{Savchenkov2004} exhibit the highest Q factors in the near-IR and visible spectral ranges (along with the highest reported optical finesse of $\mathcal{F}$ $\sim$10$^{7}$ \cite{Savchenkov2007}). Ultra-high Q factors ($>$10$^{8}$) are routinely obtained in magnesium fluoride ($\mathrm{MgF_2}$) and calcium fluoride ($\mathrm{CaF_2}$) \cite{Grudinin2006,Hofer2010} and can be as high as 10$^{11}$ in $\mathrm{CaF_2}$ \cite{Savchenkov2007}. Recently, soft fluoride crystals such as barium fluoride ($\mathrm{BaF_2}$) and strontium fluoride ($\mathrm{SrF_2}$) have equally demonstrated ultra-high Q in the near-IR \cite{Lin2014,Henriet2015}. Such ultra-high Q factors have allowed to observe Kerr comb generation \cite{Savchenkov2004} as well as temporal dissipative soliton formation \cite{Herr2013,Liang2015b} at low (mW) power levels, and enabled low phase noise microwave generation \cite{Savchenkov2004} as well as lasers with ultra narrow-linewidth \cite{Liang2015}. However, little is known what concerns the actual properties of crystalline microresonators in the mid-IR. Using QCL, recent work have reported Q factors of $\sim$ 10$^{7}$ at 4.5 $\mu$m in $\mathrm{MgF_2}$ \cite{Savchenkov2015,Grudinin2016}, $\mathrm{CaF_2}$ \cite{Savchenkov2015,Grudinin2016} and $\mathrm{BaF_2}$ microresonators \cite{Grudinin2016}, therefore being more than two orders of magnitude lower than the near-IR values.

Here we study systematically the optical Q factors of four crystalline materials transparent in the mid-IR window of the alkaline earth metal fluoride $\mathrm{XF_2}$ family (where X $=$ Ca, Mg, Ba, Sr). Note that other combinations such as X$=$Be, Ra are water soluble or toxic and therefore not suitable. To study the Q factors in the mid-IR we developed an efficient coupling technique based on an optical tapered fiber made out of chalcogenide (ChG) glass. It enables that light from a QCL can be evanescently coupled to a crystalline microresonator via a ChG uncoated tapered fiber. We show that critical coupling \cite{Cai2000} is achieved with high ideality \cite{Spillane2003}, necessary for faithful Q factor measurements, and extending this technique for the first time to the mid-IR. We measured a critical factor of $Q_{c}$ $\sim 1 \times 10^{7}$  of the $\mathrm{MgF_2}$ microresonator, a value close to the theoretical limit of multiphonon absorption at this wavelength \cite{Grudinin2016}. Using a cavity ringdown method, we demonstrate for the first time ultra-high Q ($>$10$^{8}$) mid-IR resonances in $\mathrm{BaF_2}$ and $\mathrm{SrF_2}$  microresonators. They feature ultra-high optical quality factors of Q $\geqslant$ 1.4 $\times$ 10$^{8}$ at 4.4 $\mu$m (more than a ten-fold improvement compared to previous results), exhibiting the highest observed finesse of $\mathcal{F} \sim$ 4 $\times$ 10$^{4}$ for any cavity in the mid-IR so far. Indeed the finesse of a cavity $\mathcal{F} = \dfrac{\Delta \lambda}{\delta \lambda}$ relates its free spectral range $\Delta \lambda$ to the linewidth of its resonances $\delta \lambda$ and determines the resolution of many optical measurement methods. These losses, when expressed as mirror losses in an equivalent Fabry-P\'{e}rot cavity correspond to a mirror loss of 150 ppm at 4.4 micron.

\section{Mid-IR crystalline microresonators properties and fabrication}

We are interested in crystalline materials whose transparency window includes the region of 3-5 $\mu$m, a range where QCL are commercially available. We studied four different crystalline materials such as $\mathrm{MgF_2}$, $\mathrm{CaF_2}$, $\mathrm{BaF_2}$ and $\mathrm{SrF_2}$, which are available commercially with high purity (and used e.g. for deep UV lenses). These crystals feature ultra-high Q in the near-IR \cite{Grudinin2006,Hofer2010,Lin2014,Henriet2015} and anomalous group velocity dispersion in the mid-IR, a requirement for Kerr comb and (bright) temporal soliton formation based on the Kerr nonlinearity. However, the theoretically predicted limiting loss mechanism of the Q factor in the mid-IR results from multiphonon absorption processes \cite{Hass1977,Bendow1979,Lines1986}, involving the coupling of the incident light with fundamental molecular vibrational modes of the material, a loss mechanism not observed so far in the near-IR in such crystals. Usually this reduces the mid-IR cut-off wavelength to a much shorter value than that of the transparency window's maximum wavelength. In the multiphonon-dominated regime the attenuation is given by $ \alpha = A e^{-\beta/\lambda}$ and largely surpasses any Rayleigh scattering contribution proportional to $ \dfrac{B}{\lambda^{4}}$, where $A$, $\beta$ and $B$ are materials properties \cite{Lines1986}. Therefore the Q factor limit is given by $Q = \dfrac{2\pi n}{\alpha \lambda}$ where $n$ is the refractive index and $\lambda$ the wavelength. Figure \ref{figure1}(f) presents the dependence of the Q factor with respect to the wavelength for alkalide earth metal fluoride crystalline materials \cite{Gorodetsky1996}. It reveals that even though a material displays low Rayleigh scattering and ultra-high Q in the near-IR, it will be subject to multiphonon absorption in the mid-IR that increases by $\sim$2-6 orders of magnitude at 4.5 $\mu$m compared to 1.5 $\mu$m for these crystal families. Based on theoretical considerations alone, $\mathrm{BaF_2}$ and $\mathrm{SrF_2}$ seem ideal candidates to achieve ultra-high Q in the mid-IR in comparison to $\mathrm{MgF_2}$ due to lower multiphonon absorption. Though there has been considerable interest in highly-transparent infrared materials for the last decades, only spectrophotometric measurements were carried out to study absorption due to technical limitations of using microresonators in the mid-IR \cite{Hass1977,Bendow1979,Lines1986}. We emphasize that conventional methods (e.g. Fourier transform infrared spectroscopy) to investigate loss are not suitable to probe the losses at the level of ppm, as required to achieve ultra-high Q. 

We fabricated our microresonators either from disk or cylinder blanks. The microresonators were first shaped by grinding or diamond-cutting tool and polished in an air-bearing spindle by successive smaller diamond particle slurries in order to obtain a smooth protrusion as visualized in Fig.\ref{figure1}b. Their final diameters are $\sim 5$~mm. We analyzed the microresonators in a scanning electron microscope. For instance we determined transverse radii of the crystalline $\mathrm{MgF_2}$  disk of $\lesssim$ 50 $\mu$m. The intensity profile of the fundamental WGM of the $\mathrm{MgF_2}$  microresonator in the mid-IR is displayed in Fig.\ref{figure1}(c). From finite element model simulations we obtained an effective mode area of $A_{\mathrm{eff}}\sim 600  \ \mu$m$^{2}$ at $\lambda \sim$ 4.4$\ \mu$m. To ensure that we are not limited by scattering losses on residual surface roughness, their optical quality factors were first measured at $\lambda = 1.55 \ \mu$m with a tunable, narrow-linewidth (short-term $<100$~kHz) fiber laser using silica tapered fibers to excite the WGM \cite{Hofer2010}. The $\mathrm{MgF_2}$ disk features loaded optical factors of Q $\geqslant$ 5$\times$ 10$^{8}$; $\mathrm{CaF_2}$ and $\mathrm{SrF_2}$ exhibit Q factors of $\sim$ 2 $\times$ 10$^{9}$ and the $\mathrm{BaF_2}$ cylinder features Q $\sim$ 6.4 $\times$ 10$^{9}$ (detailed in a further section of the manuscript and in Figure \ref{figure5}). 

 \begin{figure}[ht]
		\includegraphics[width=0.95\columnwidth]{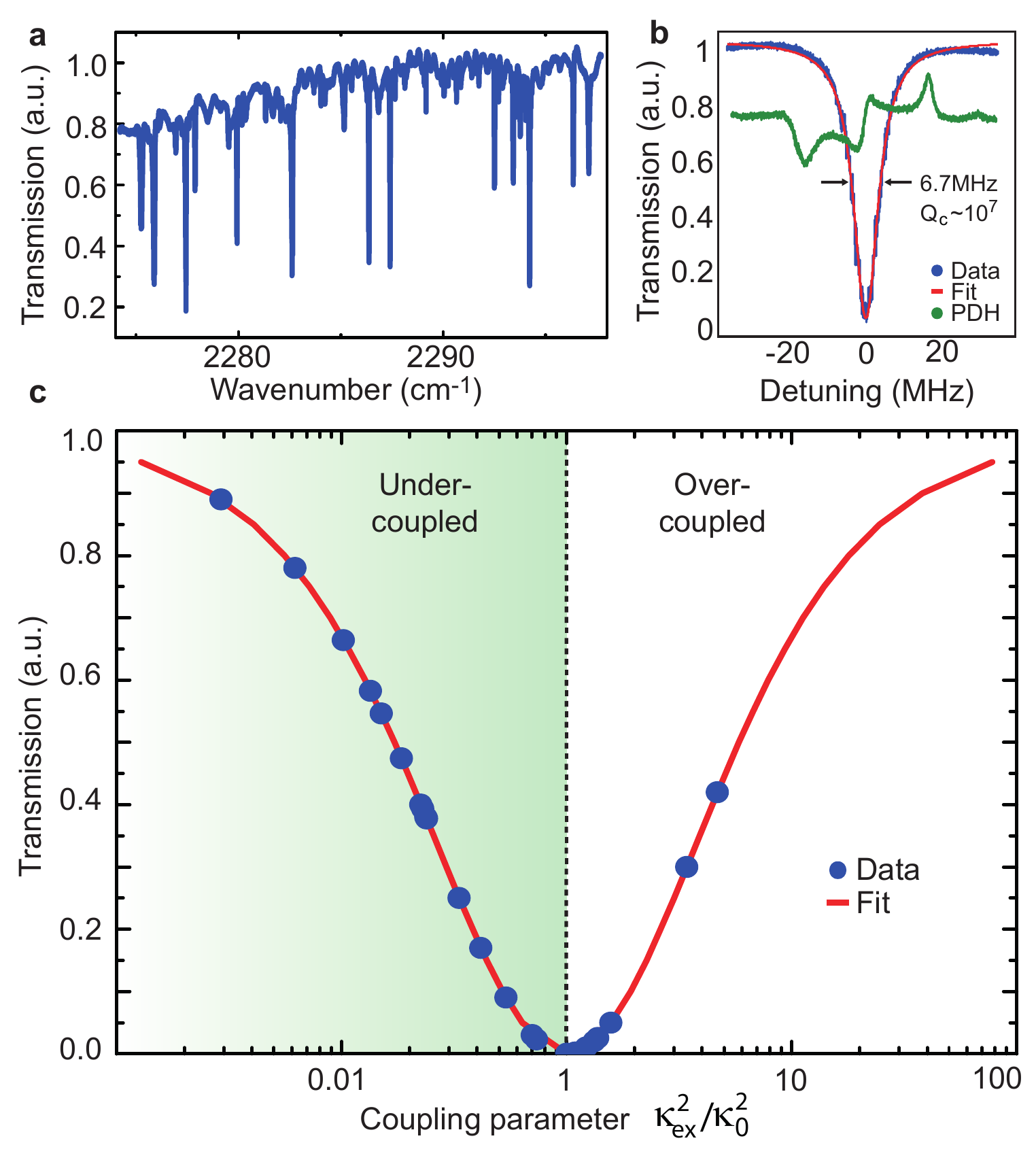}
	\caption{\textbf{Characterization of taper-resonator coupling efficiency in the mid-IR spectral region around 4.5~$\mu$m.} \textbf{a} Representative
		transmission spectrum composed of several resonance families over a wide mid-IR range when taper-resonator coupling is achieved. \textbf{b} Measurement of a resonance linewidth at critical coupling with frequency calibration provided by a Pound-Drever-Hall (PDH) signal (green curve) and Lorentzian fit (red curve). The typical FWHM width of $\kappa /2 \pi$ $\sim 6.7$~MHz corresponds to a critically coupled quality factor of $Q_{c}\sim 1.0 \times 10^{7} $. \textbf{c} Transmission as a function of the coupling parameter $\kappa_{\mathrm{ex}}^{2}/ \kappa_{0}^{2}$ for varying taper waist radius. The dashed line marks the critical coupling point ($\kappa_{ex}=\kappa_{0}$). The experimental data (blue circles) are consistent with the theoretical fit (red curve) demonstrating that the ChG taper behaves as a nearly ideal coupler in mid-IR with close to unity ideality.}
	\label{figure2}
\end{figure}

\newpage
\begin{figure*}[htbp]
  	  	\includegraphics[width=\linewidth]{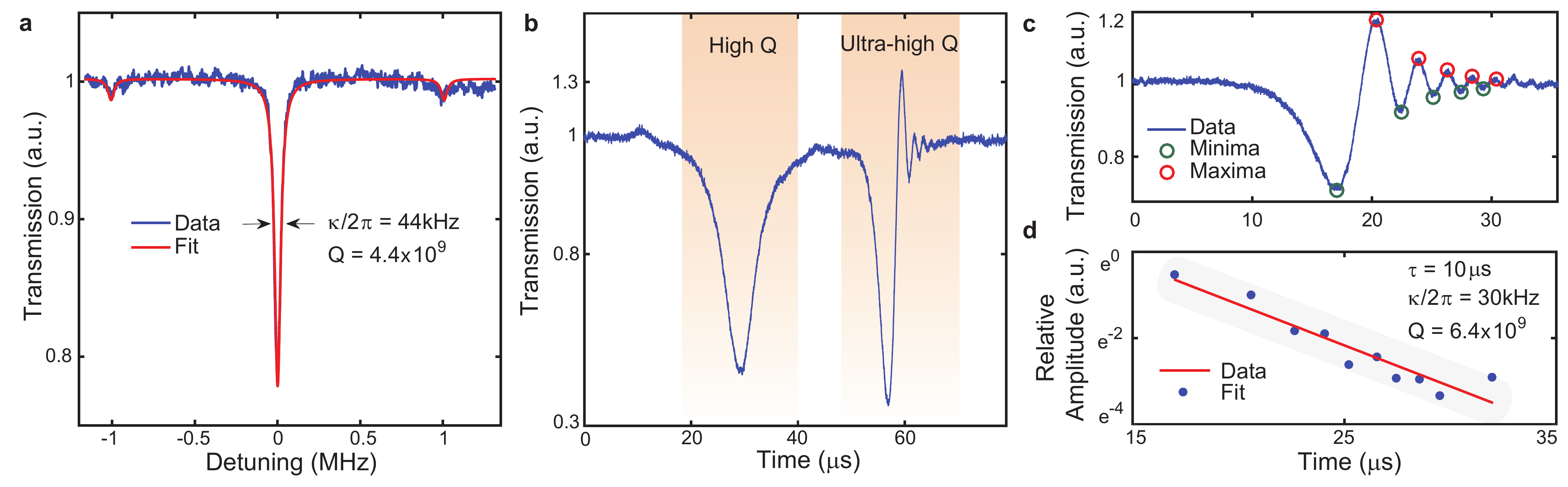}
  	\caption{\textbf{Near-IR characterization of a $\mathrm{BaF_2}$ microresonator. a} Linewidth measurement of an under-coupled resonance with calibration sidebands (resulting from phase modulation of the laser) at 1 MHz and Lorentzian fit (red line). We extract a typical FWHM of $\kappa /2 \pi$ $\sim$ 44 kHz resulting in an optical factor of $Q\sim 4.4 \times 10^{9} $. \textbf{b} Cavity-ringdown measurements. We observed the transmission spectrum of two resonances while scanning at the same speed. Only the ultra-high Q mode features a ringdown signal (right resonance). \textbf{c} Analysis of ringdown structure. We extract successive amplitudes of maxima (red circles) and of minima (green circles) to fit the evolution of the relative amplitude ringdown \textbf{d} .  Theoretical fit of the ringdown relative amplitude (red line). The measured amplitude decay of $\tau$ = 10 $\mu$s results in a linewidth of $\kappa /2 \pi$ = 30 kHz and a Q factor of 6.4 $\times$ 10$^{9}$. }
    \label{figure3}
\end{figure*}
\begin{figure*}[htbp]
  	  	\includegraphics[width=\linewidth]{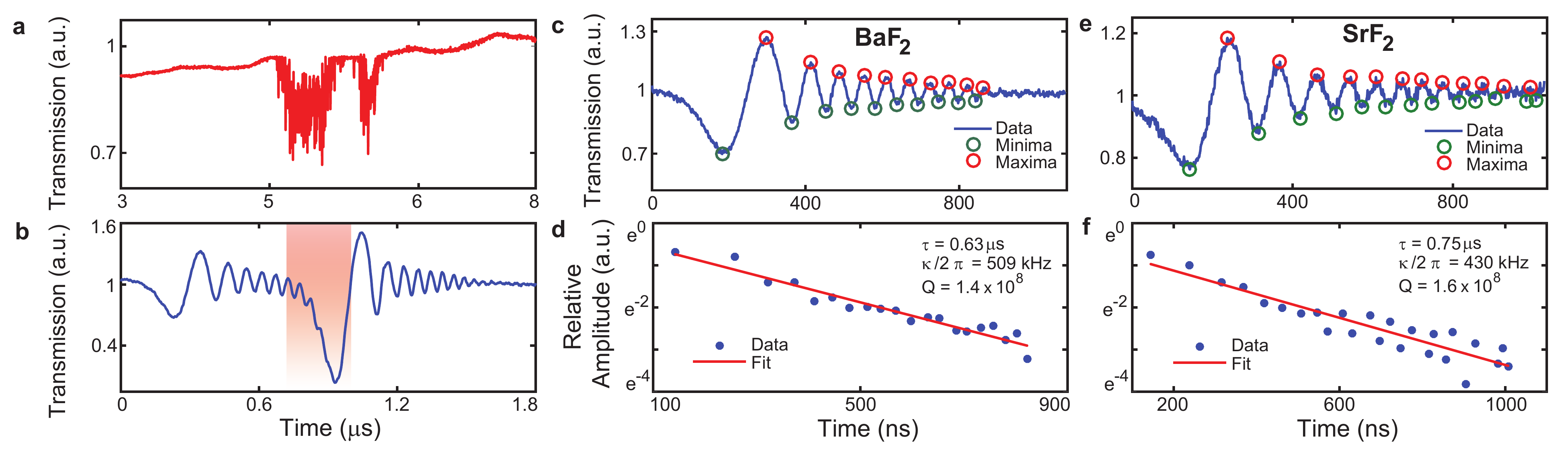}
  	\caption{\textbf{Mid-IR swept frequency ringdown characterization of $\mathrm{BaF_2}$ and $\mathrm{SrF_2}$ microresonators}. \textbf{a} Thermal distortions of a resonance in the $\mathrm{BaF_2}$ crystalline microresonator. Thermal-optical dynamics prevent measuring the linewidth and the transmission accurately using standard frequency modulation spectroscopy method. \textbf{b} Using a cavity swept laser ringdown method, we observe modal coupling between two ultra-high Q resonances. \textbf{c,e} Transmission spectra of the ringdown signal of an ultra-high Q resonance in mid-IR at 4.4 $\mu$m using $\mathrm{BaF_2}$ (c) and $\mathrm{SrF_2}$ (e) \textbf{d,f} Exponential fits (red lines) of the measured data (blue circles) extracted from (c,e). In $\mathrm{BaF_2}$ (d), the measured amplitude lifetime of $\tau$ = 0.63 $\mu$s results in a linewidth of $\kappa /2 \pi$ = 509 kHz and a corresponding Q factor of $\sim$1.4 $\times$ 10$^{8}$. In $\mathrm{SrF_2}$ (f), the measured amplitude lifetime of $\tau$ = 0.75 $\mu$s results in a linewidth of $\kappa /2 \pi$ = 430 kHz and a corresponding Q factor of $\sim$1.6 $\times$ 10$^{8}$.}
  	\label{figure4}
  \end{figure*}
\newpage

\section{Uncoated chalcogenide tapered fibers as high ideality mid IR couplers}

To measure far in the infrared, we used a mode-hop-free tunable QCL (Daylight Solutions Inc.). To couple QCL light into the microresonator we developed optical tapered fibers made out of ChG glass. ChG fibers are particularly attractive due to their low loss in the mid-IR \cite{Eggleton2011} and commercial availability. However they were never used previously as a low loss, uncoated tapered optical waveguide because of technical fabrication challenges in their tapering process and maintenance, due to the low melting point of ChG glasses ($\sim 200^{\circ}$C for As$_{2}$S$_{3}$). We have successfully developed a tapering setup with a feedback controlled electrical heater to adiabatically pull low loss uncoated tapered ChG fibers. To verify the theoretically expected taper waist, scanning electron microscope images were taken to optimize the tapering process.
The ChG (As$_{2}$S$_{3}$) tapered fiber was fabricated from a commercially available, IRflex IRF-S 10 nonlinear mid-IR fiber, made from extra high purity ChG glass. The fiber is transparent from 1.5 to $6.5 \  \mu$m and has a core diameter of $10 \  \mu$m and a hig nonlinear refractive index of $n_{2}=2.7$.  
The ChG tapered fiber was pulled down to a diameter to phase match the fundamental WGM. For instance, for a MgF$_{2}$ microresonator radius of $R\sim 2.5$~mm, the optimum taper waist in the mid-IR corresponds to a subwavelength diameter of 1.2 $\mu$m (Fig.\ref{figure1}(d)). The experimental setup is described in Fig.\ref{figure1}e. The pump laser is a 200 mW continuous wave QCL external cavity laser, tunable from 4.385 to 4.58 $\mu$m. Tapered fiber and microresonator are kept under a dry and inert atmosphere to preserve from degradation. The QCL light is evanescently coupled to the crystalline microresonator using the ChG tapered fiber. When the taper-resonator coupling is achieved, several resonance families are observed at $\lambda \sim 4.4 \ \mu$m, as illustrated in Fig.\ref{figure2}(a).

A unique property of taper-coupled microresonators is that phase matching (and thus the coupling) can be finely tuned by translating the microresonator along the taper relative to the waist \cite{Cai2000}. We investigated in detail the mid-IR coupling behavior between the ChG taper and the $\mathrm{MgF_2}$ crystalline microresonator for a typical mode family. For an unity ideality coupler \cite{Spillane2003}, the coupling parameter is given by $\kappa_{\mathrm{ex}}^{2}/ \kappa_{0}^{2}=(1 \pm \sqrt{T})/(1 \mp \sqrt{T}) $, where $\kappa_{0}$ is the intrinsic loss rate, $\kappa_{\mathrm{ex}}$ the photon loss rate due to coupling to the microresonator and $T$ the transmission on resonance. The upper signs are used for transmission values $T$ in the over-coupled regime ($\kappa_{0} < \kappa_{\mathrm{ex}} $) and the lower signs for the under-coupled regime ($\kappa_{0} > \kappa_{\mathrm{ex}} $). We recorded the transmission spectrum while scanning the laser over resonance for different taper waist radii. We measured the corresponding full-width at half maximum (FWHM) linewidth by setting up a Pound-Drever-Hall (PDH) technique with a mid-IR electro-optic phase modulator (Qubig GmbH) to provide frequency calibration. Figure \ref{figure2}(c) shows the normalized transmission as a function of the parameter $\kappa_{\mathrm{ext}}^{2}/ \kappa_{0}^{2} $. Critical ($\kappa_{0}= \kappa_{\mathrm{ex}} $) and strong overcoupling up to $\kappa_{\mathrm{ex}}/ \kappa_{0}\sim 6 $ are observed for optimum taper diameters. We emphasize that this represents the first achievement of critical coupling in crystalline microresonators in the mid-IR region. This reveals that the ChG taper behaves as a nearly ideal coupler, with close to unity ideality \cite{Spillane2003}. We measured a linewidth of $\kappa / 2\pi$ $\sim 6.7$~MHz at critical coupling (red curve in Fig.\ref{figure2}b.) yielding an optical factor of $Q_{\mathrm{c}}$ $\sim 1.01 \times 10^{7}$ for $\mathrm{MgF_2}$. Our measurements reveal that uncoated ChG tapers can thus extend the efficient tapered fiber coupling method deep to the mid-IR regime. It enables a complete characterization control and use of microresonators in the mid-IR, in contrast to other methods such as prism coupling \cite{Savchenkov2015,Grudinin2016}.

\section{Near- and mid-IR cavity ringdown measurements of crystalline microresonators}

After $\mathrm{MgF_2}$ characterization, we studied other fluoride crystals employed in this work ($\mathrm{CaF_2}$, $\mathrm{BaF_2}$, $\mathrm{SrF_2}$) and first characterized them in the near-IR setup. We measured the optical Q factor using both linewidth calibration and a swept laser cavity ringdown technique. The cavity ringdown method enables a measurement of the quality factor independently of the thermal nonlinearity (hence at higher pumping powers) and of the laser linewidth. Fig.\ref{figure4}(a) shows a typical resonance obtained in the under-coupled regime for a pump power of P $\leqslant$ 1 mW in $\mathrm{BaF_2}$. The Lorentzian fit gives a FWHM linewidth of $\kappa / 2\pi$ $\sim$ 44 kHz at 1555 nm, resulting in an optical factor of Q $\sim$ 4.4 $\times$ 10$^{9}$. From the coupling curve we can extract moreover an intrinsic linewidth of $\kappa_{0}/2 \pi \sim$ 22 kHz (corresponding to an intrinsic Q of 0.9$\times$ 10$^{10}$). In addition, we performed swept laser cavity ringdown measurements \cite{Hofer2010} to first calibrate the method in the near-IR. When an ultra-high Q resonator is excited with a laser whose frequency is linearly swept across the resonance with a duration shorter than the cavity lifetime, its transmission spectrum shows oscillations (see Fig\ref{figure4}.(b)). These oscillations result from the beating of the transiently build-up light inside the resonator that decays into the fiber, with that of the swept laser source. Importantly, when the laser is swept fast, the two components will have a different beat frequency giving rise to an exponentially decaying oscillation. In the case of a lower Q mode (with negligible cavity buildup), no ringdown signal is observed at the same scan speed, as highlighted in Fig.\ref{figure4}.(b). We analyzed the transmission spectrum of the ringdown structure in Fig\ref{figure4}.(c) by measuring the amplitude decay  $\tau$ of the remitted light \cite{Savchenkov2007}. Its theoretical fit (Fig\ref{figure4}(d)) gives a measured amplitude decay $\tau$ of 10 $\mu$s, corresponding to an intrinsic cavity Q$_{0}$ of $\sim$ 6.4 $\times$ 10$^{9}$. This value is close to the one derived with the frequency modulation spectroscopy, corroborating the faithfullness of the method.  In $\mathrm{CaF_2}$ and $\mathrm{SrF_2}$ microresonators we measured $\tau \sim$ 3 $\mu$s, corresponding to an intrinsic cavity Q$_{0} \sim$ 1.8 $\times$ 10$^{9}$, in agreement with the values derived from frequency modulation spectroscopy. We precise that we did not resort to annealing and/or baking procedures to improve Q factors of our microresonators \cite{Savchenkov2007}. 
 
Having established the ultra-high Q resonance in the near-IR, and having developed the mid-IR ChG tapered fiber method, we afterwards measured Q factors in the mid-IR. Phase matching between the ChG tapered fiber and the microresonator is achieved by translating the taper position. In contrast to the near-IR, when coupling is achieved, we observed large thermal instability in the mid-IR \cite{Fomin2005}. These thermal instabilities induce large distortions of the resonance (see Fig.\ref{figure4}.(a)) preventing any reliable measurement of its linewidth using the PDH error signal as a calibration. In that case, the optical Q factor can only be inferred by cavity ringdown measurements. We scanned the QCL using its laser current modulation in order to observe the cavity ringdown signal. Amplitude and cavity lifetime depend on coupling and linewidth of the resonance. When two resonances are really close, we observe a modal coupling between the two oscillatory decaying signals (see Fig.\ref{figure4}.(b)) \cite{Trebaol2010}. The $\mathrm{BaF_2}$ transmission spectrum of a typical ringdown structure is displayed in Fig.\ref{figure4}.(c). By measuring the amplitude decay $\tau$ of the remitted light (Fig.\ref{figure4}(d)), we obtained a $\mathrm{BaF_2}$ intrinsic factor of Q$_{0}=\omega \tau /2\sim$ 1.4 $\times$ 10$^{8}$. 

\begin{table}[h]
\centering
\caption{\bf Experimentally measured mid-IR absorption properties of different crystalline fluoride materials, at 4.4 $\mu$m }
\begin{tabular}{ccccc}
\hline
Crystal & $\kappa / 2\pi$ & $Q_{0}$ &  $\alpha$ (cm$^{-1}$)&$\mathcal{F}$\\
\hline
$\mathrm{MgF_2}$ & 3.4 MHz & 2 $\times$ 10$^{7}$ & 9 $\times$ 10$^{-4}$&4 $\times$ 10$^{3}$\\
$\mathrm{CaF_2}$ & 710 kHz & 9.6 $\times$ 10$^{7}$ & 2 $\times$ 10$^{-4}$&2 $\times$ 10$^{4}$\\
$\mathrm{BaF_2}$ & 509 kHz & 1.4 $\times$ 10$^{8}$ & 1.4 $\times$ 10$^{-4}$&3 $\times$ 10$^{4}$\\
$\mathrm{SrF_2}$ & 430 kHz & 1.6 $\times$ 10$^{8}$ & 1.2 $\times$ 10$^{-4}$&4 $\times$ 10$^{4}$\\
\hline
\end{tabular}
  \label{tab:table}
\end{table}

\begin{figure}[h]
	\centering
	\includegraphics[width=\columnwidth]{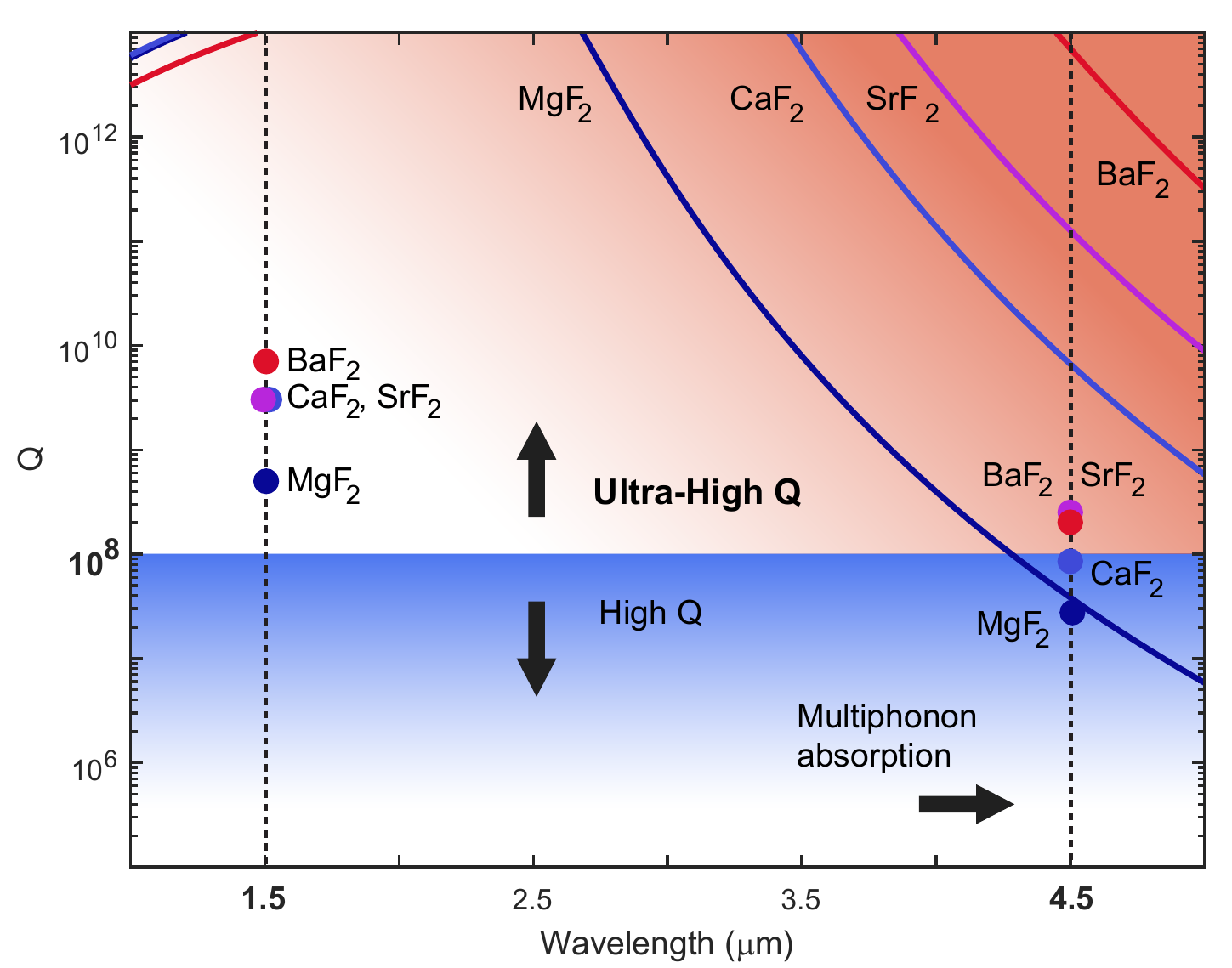}
	\caption{\textbf{Mid-IR ultra-high Q microcavities based on alkaline earth metal fluoride crystalline materials.} Measurements for different fluoride crystals of the $\mathrm{XF_2}$ family (where X $=$ Ca, Mg, Ba, Sr) prove the possibility of attaining the ultra-high Q regime in the mid-IR. Except for $\mathrm{MgF_2}$, for which we reach the theoretical limit imposed by multiphonon absorption at room temperature, other materials offer $Q \geqslant 10^{8}$ around 4.5 $\mu$m. The lines represent the theoretical multiphonon absorption limit of Q with respect to the wavelength. The circles represent our experimental values. Despite the clear differences with the near-IR region, mid-IR cavities are able to overcome the high-Q regime achieving Q$>$10$^{8}$.
		}
	\label{figure5}
\end{figure}

$\mathrm{CaF_2}$ and $\mathrm{SrF_2}$ crystals were studied by applying the same methods. The ringdown analysis in the case of $\mathrm{SrF_2}$ is presented in Fig.\ref{figure4}(e). We measured an amplitude decay time $\tau$ of 0.75 $\mu$s (Fig.\ref{figure4}(f)), resulting in an intrinsic factor of Q$_{0}=\omega \tau /2\sim$ 1.6 $\times$ 10$^{8}$. We precise that no ringdown were observed when measuring the Q factor of the $\mathrm{MgF_2}$ microresonator under the same conditions, confirming that $\mathrm{MgF_2}$ does not feature ultra-high Q factors in the mid-IR \cite{Grudinin2016}. Table \ref{tab:table} and Fig.\ref{figure5} summarize the results of our measurements. From these measurements we extracted the limiting values of the mid-IR intrinsic absorption $\alpha \leqslant \dfrac{2\pi n}{Q \lambda}$ at 4.4 $\mu$m. We show that the absorption can thus be as low as 100 ppm/cm for $\mathrm{SrF_2}$. Translated to a single round-trip in a Fabry-P\'{e}rot cavity of length $L/2$ (where $L=2\pi r$ for a WGM of radius $r$) and finesse $\mathcal{F}$, this value amounts to optical losses (absorbance) of the order of $\alpha L=\dfrac{2\pi}{\mathcal{F}}$, i.e. 150 ppm.

\section{Conclusions}
In summary, we have demonstrated for the first time mid-IR ultra-high Q crystalline resonators made from commercially available $\mathrm{BaF_2}$ and $\mathrm{SrF_2}$ crystals at 4.5 micron, a region of high interest due to the transparency of the atmosphere and absorption of toxic or greenhouse molecules (e.g. CO, $\mathrm{CO}_2$, $\mathrm{SO}_2$ or $\mathrm{CO}_3$) and the availability of QCL laser sources. Moreover we show that $\mathrm{MgF_2}$ crystals operate close to the fundamental limit imposed by multiphoton absorption in the mid-IR. We show that uncoated chalcogenide tapered fibers can be ideal and efficient couplers deep to the mid-IR. Together with cavity ringdown methods, our platform enables precise measurements of quality factor, overcoming previous limitations. 
We show that a finesse as high as $\mathcal{F} \sim$ 4 $\times$ 10$^{4}$ is achievable with $\mathrm{BaF_2}$ and $\mathrm{SrF_2}$ in the mid-IR for cavities as small as few millimeters in diameter. This finesse represents a more than 1 order of magnitude improvement over prior high finesse cavities in this wavelength range \cite{Schwarzl2007,AlligoodDeprince2013,Foltynowicz2013}.  Our results in the mid-IR pave the way to the next generation of ultra-stable sources and ultra-precise spectrometers in the molecular fingerprint region and can further leverage QCL technology, by e.g. enabling injection locked QCL similar to technology developed in the near-IR \cite{Liang2015,Liang2015b}. Despite the differences with near-IR, we prove that the mid-IR region is not limited to the high-Q regime when proper materials are used. The materials crystalline nature leads to low thermodynamical noise \cite{Cole2013} and moderate multiphonon absorption in the highly-relevant mid-IR region. 
In addition, combining QCL with mid-IR ultra-high Q crystalline microresonators opens a route for mid-IR Kerr comb or soliton generation \cite{Herr2013,Yi2015,Brasch2016}.

\section*{Funding Information}
This work was gratefully supported by the Defense Advanced Research Program Agency (DARPA) under the PULSE program and by grant N$^{\circ}$ IZLRZ2$\_$163864 from the Swiss National Science Foundation (SNSF). C. L. gratefully acknowledges support from the European Commission through Marie Skodowska-Curie Fellowships: IEF project 629649.

\section*{Acknowledgments}

We gratefully acknowledge helpful discussions with Yanne Chembo. We acknowledge technical support from Arnaud Magrez, Qubig GmbH and IRflex Corporation.

\vspace{\fill}

\renewcommand{\emph}{\textit}
\bibliographystyle{naturemag}
\bibliography{carobib}

\end{document}